\documentclass[11pt a4paper]{article}
\usepackage{jheppub}
\usepackage{bm}

\newcommand{\be}{\begin{equation}}
\newcommand{\dd}{\displaystyle}
\newcommand{\ee}{\end{equation}}
\newcommand{\bea}{\begin{eqnarray}}
\newcommand{\eea}{\end{eqnarray}}

\newcommand{\nn}{\nonumber}
\newcommand{\de}{\partial}
\def\nn{\nonumber}
\def\de{\partial}

 \def\slash#1{\setbox0=\hbox{$#1$}#1\hskip-\wd0\dimen0=5pt\advance
       \dimen0 by-\ht0\advance\dimen0 by\dp0\lower0.5\dimen0\hbox
         to\wd0{\hss\sl/\/\hss}}
\def\be{\begin{equation}}
\def\dd{\displaystyle}
\def\ee{\end{equation}}

\def\bea{\begin{eqnarray}}
\def\eea{\end{eqnarray}}
\def\7{\tilde}
\def\8{\hat}

 \def\slash#1{\setbox0=\hbox{$#1$}#1\hskip-\wd0\dimen0=5pt\advance
       \dimen0 by-\ht0\advance\dimen0 by\dp0\lower0.5\dimen0\hbox
         to\wd0{\hss\sl/\/\hss}}

\def\alp{{\mu}}

\def\txi{\tilde\xi}

\def\tZ{\tilde Z}
\def\tc{\tilde c}
\def\tepsilon{\tilde\epsilon}
\def\hi{\hat i}
\def\hj{\hat j}

\begin{document}

\subheader{\hfill {\rm ICCUB-17-017}}

\title{Non-relativistic Spinning Particle in a Newton-Cartan Background}

\author[a]{Andrea Barducci}
\affiliation[a]{Department of Physics and Astronomy, University of Florence and
INFN, Via G. Sansone 1, 50019 Sesto Fiorentino (FI), Italy}

\author[a]{Roberto Casalbuoni}
\author[b]{Joaquim Gomis}\affiliation[b]{Instituto de Ciencias F\'{\i}sicas y Matem\'aticas,
Universidad Austral de Chile,\\  Campus Isla Teja,Valdivia, Chile.}
\affiliation{Centro de Estudios Cient\'{\i}ficos (CECs), Av. Arturo Prat 514, Valdivia 5110466, Chile.}
\affiliation{Departament de F\'{\i}sica Qu\`antica i Astrof\'{\i}sica and Institut de Ci\`encies del Cosmos (ICCUB), 
Universitat de Barcelona, Mart\'{\i} i Franqu\`es 1, E-08028 Barcelona,
 Spain}

\emailAdd{barducci@fi.infn.it}\emailAdd{casalbuoni@fi.infn.it}\emailAdd{gomis@ecm.ub.es} 

\begin{abstract}
{
We construct the action of a non-relativistic spinning particle moving in a general torsionless Newton-Cartan background. The particle does not follow the geodesic equations, instead the motion is governed by the non-relativistic analog of 
Papapetrou equation.  The spinning particle is described in terms of Grassmann variables. In the flat case the action is invariant under the non-relativistic analog
of space-time vector supersymmetry.
}

\end{abstract}
\maketitle

\section{Introduction}\label{sec:0}
{
Strong correlated systems in condensed matter  have found in non-relativistic holography a new technique to understand their behaviour, see for example
\cite{review,review1}.
The reason is that holography is a strong-weak duality mapping. If, in the screen we have a strong coupled quantum field theory, in the bulk we have a weak description of string theory. If we are in a situation  where the curvature of the space time is small, we can use classical gravity
instead of full string theory. 
In the case of non-relativistic holography in the bulk one can use an Einstein metric with non-relativistic isometries \cite{Son:2008ye,Balasubramanian:2008dm,Herzog:2008wg,Kachru:2008yh}
or   non-relativistic gravities in the bulk \cite{Son:2013rqa,Geracie:2016bkg,Janiszewski:2012nb,Wu:2014dha},
like Newton-Cartan gravity \cite{Cartan:1923zea} or Horava gravity \cite{Horava:2009uw}.
Having in mind this picture, it is interesting to study matter coupled to non-relativistic gravity. For example particles  \cite{Kuchar:1980tw}, %
and 
extended objects \cite{Andringa:2012uz}  and 
Galilean field theories \cite{Jensen:2014aia,Hartong:2014pma} coupled to a Newton-Cartan background.

In this paper we construct the action of a non-relativistic spinning particle moving in a general torsionless Newton-Cartan background. The particle does not follow the geodesic equations, instead the motion is governed by the non-relativistic analog of  the
Papapetrou equation \cite{Papapetrou:1951pa}.  
The spinning particle is described in terms of Grassmann variables. In the flat case the action is invariant under the non-relativistic analog
of  space-time vector supersymmetry, called VSUSY  \cite{Casalbuoni:2008iy}. This model is obtained from the relativistic
spinning particle \cite{Barducci:1976qu} with variables $\xi_\mu, \xi_5$. 

In the flat case, the limit is done at the level of the coordinates of the
particle, the form of the limit being  suggested by the contraction of the algebra of VSUSY
to a non-relativistic version, that we will call NR-VSUSY. 
{
The model is  invariant under this non-relativistic symmetry  and 
also invariant under diffeomorphisms and the non-relativistic VSUSY version of kappa-symmetry \cite{deAzcarraga:1982dw,Siegel:1983hh}. 
The associated two first class constraints give rise to the non-relativistic mass-shell constraint and to a Levy-Leblond type of constraint \cite{LevyLeblond:1967zz,Gomis:1985uf}.

In order to get the non-relativistic spinning particle in a torsionless Newton-Cartan background, our starting point is a relativistic spinning particle coupled to a general Einstein background \cite{Barducci:1976wc} and  to a U(1) gauge field with vanishing field strength \cite{Bergshoeff:2015uaa,Gomis:2000bd}.
In this case  the non-relativistic limit is done on the background fields and not on the coordinates.
}
{
We find that  the first class character of the constraints imposes the  condition that the $U(1)$ connection surviving in the non-relativistic limit  must have zero field strength.}

The  paper is organized as follows: in Section II we perform a contraction of the VSUSY algebra leading to its non-relativistic version. In Section III we introduce the action of the VSUSY particle \cite{Casalbuoni:2008iy}. Then,  we define the non-relativistic limit of this model by performing a  transformation of the dynamical variables in agreement with the results of Section II. In Section IV we study the equations of motion of the non-relativistic model showing the presence of two first-class constraints that are associated to the diffeomorphism invariance of the model, and to the non-relativistic version of the kappa-symmetry owned by the relativistic model. This world-line symmetry is investigated in Section V. In Section VI we start again from the relativistic model coupled to a general four-dimensional background metric  \cite{Barducci:1976wc} and define the limit to a torsionless Newton-Cartan metric, using the group contraction defined in Section II. 
 In Section VII we derive the equations of motion, showing that, as in the relativistic case, the geodesic equations are corrected by a term proportional to the spin of the particle coupled to a 
 Newton-Cartan curvature.
In Section VIII our conclusions and an outlook.}

\section{Algebra contraction}
\label{sec:2}

{
Many dynamical models can be obtained as non-linear realizations of a space-time symmetry group, $G$. Examples are the relativistic point particle \cite{Gauntlett:1989qe}, the relativistic spinning particle \cite{Casalbuoni:2008iy}, the D-branes \cite{Bagger:1996wp} etc. An interesting question is what happens to these models  if we consider a contraction of  the Lie algebra of $G$, Lie-$G$. Another related question arises if one couples the original model to a gravitational field. Precisely one can ask what happens to the gravitational field after the contraction. This last question will be discussed later on.
To be more explicit let us define the contraction of a given algebra. Suppose that our starting algebra  {
(or a superalgebra)} satisfies the commutation relations
\be
[X_\alpha, X_\beta] =f_{\alpha\beta}^\gamma X_\gamma
\ee
and let us define an invertible linear transformation depending on a parameter $\omega$
\be
Y_\alpha =\sum_\beta A_\alpha^\beta (\omega)X_\beta.
\ee
The Lie algebra satisfied by the new generators will be
\be
[Y_\alpha, Y_\beta] ={\bar f}_{\alpha\beta}^{\,\gamma}(\omega) Y_\gamma,
\ee
with
\be
{\bar f}_{\alpha\beta}^{\,\gamma}(\omega)= A_\alpha^\sigma(\omega)A_\beta^\tau(\omega)
  (A^{-1})_\delta^\gamma (\omega)f_{\sigma\tau}^\delta.
  \ee
  Then, consider the limit $\omega\to \infty$ and suppose that
the limit of the new structure constants ${\bar f}_{\alpha\beta}^{\,\gamma}(\omega)$ is finite. 
When the limit is non-singular,
we  say that   the algebra of the  $Y_\alpha$'s is a "contraction" of the algebra of the $X_\alpha$'s. Notice that  the  contracted algebra is not   equivalent to the original one. We will now define a non-relativistic contraction of the relativistic VSUSY algebra to the non-relativistic version, NR-SUSY.

}

The VSUSY algebra is a graded extension of the Poincar\'e algebra, with odd generators $G_\mu$ and $G_5$, 
the relevant brackets being (see \cite{Casalbuoni:2008iy})

\be [M_{\mu\nu},M_{\rho\sigma}]=-i\eta_{\nu\rho}M_{\mu\sigma}-
i\eta_{\mu\sigma}M_{\nu\rho}+i\eta_{\nu\sigma}M_{\mu\rho}+
i\eta_{\mu\rho}M_{\nu\sigma},\label{eq:ext0} \ee \be
[M_{\mu\nu},P_\rho]=i\eta_{\mu\rho}P_\nu-i\eta_{\nu\rho}P_\mu,\qquad
[M_{\mu\nu},G_\rho]=i\eta_{\mu\rho}G_\nu-i\eta_{\nu\rho}G_\mu,
\label{eq:ext01} \ee \be [G_\mu,G_\nu ]_+=\eta_{\mu\nu}Z,~~~[G_5,G_5
]_+= Z_5 \label{eq:ext1},\ee \be
[G_\mu,G_5]_+=-P_\mu\label{eq:ext2}, \ee
where the bracket $[\cdot,\cdot]_+$ defines, as usual, the anticommutator.
This algebra involves also two scalar central charges  $Z$ and $Z_5$
Here we make use of the flat  metric 
\be\eta_{\mu\nu}=(-,+,+,+).\ee
 As we will see, it is useful to introduce the following combination of the central charges
\be
Z_\pm=Z\pm  Z_5.
\label{eq:6}
\ee

In order to define the contraction we introduce a dimensionless parameter $\omega$ to be sent to infinity. .
Then, we relate the relativistic generators to the non-relativistic ones, by the following equations
\bea
&\displaystyle{P_0=\frac{\alpha}{\omega} H-\omega\tZ_-,~~~Z_-=-\frac{2(1+\alpha)}\omega H  + 2\omega \tZ_-,~~~Z_+=2\omega\tilde Z_+},&\nn\\&
\displaystyle{G_0=\frac 1{\sqrt{2\omega}}Q_++ \sqrt{\frac{\omega^3}2} Q_-,~~~G_5=\frac 1{\sqrt{2\omega}}Q_+- 
{\sqrt{\frac{\omega^3}2}} Q_-,~~~ G_i=\sqrt{\frac\omega 2} Q_i},&\nn\\&
\ M_{0i} = \omega B_i,&
\label{eq:8}
\eea
whereas all the other variables are left unchanged and $\alpha$ is a 
 dimensionless parameter.
 We will need also the inverse relations. 
\bea
&\displaystyle{H=\frac\omega 2\left(-2P_0-Z_-\right),~~~~\tZ_- =-\frac 1{2\omega} \left(\alpha Z_-+2(1+\alpha)P_0\right),~~~
\tilde Z_+=\frac 1{2\omega} Z_+},&\nn\\&
\displaystyle{Q_+ =\sqrt{\frac\omega 2}\left(G_0+G_5\right),~~~~Q_- =\sqrt{\frac 1{2\omega^3} }\left(G_0-G_5\right),~~~~
Q_i=\sqrt{\frac 2\omega}G_i},\nn\\&
\displaystyle{B_i=\frac 1\omega M_{0i}}.&
\label{eq:9}
\eea
The  relevant 
commutators and anticommutators, in the limit $\omega\to\infty$,  are
\be
[B_i,H]=iP_i,~~~[B_i,P_j]=i\delta_{ij} \tZ_-,\ee
\bea
&[Q_+,Q_+]_+= H,~~~[Q_-,Q_-]_+ = 0,~~~[Q_+,Q_-]_+=-\tilde Z_+,&\nonumber\\ & [Q_i,Q_j]_+=2\delta_{ij}(\tZ_++\tZ_-),~~~[Q_+,Q_i]_+=-P_i,&\label{eq:11}\eea
\be
[B_i,Q_+]= -\frac i2 Q_i,~~~[B_i,Q_j]=-i\delta_{ij}Q_-,~~~[B_i,Q_-]=0.\ee

If we put $Q_-=0, \tZ_+=0$, the previous algebra collapses to the algebra found
in \cite{Gauntlett:1990nk}.
To complete the analysis of the 
NR-VSUSY algebra we will consider the relativistic quadratic Casimir \cite{Casalbuoni:2008ez,Casalbuoni:2009en}:
 $P^2-Z Z_5$.
  We use  $Z=(Z_++Z_-)/2$, $Z_5=(Z_+-Z_-)/2$ (see eq. (\ref{eq:6})). Expanding in powers of $ \omega$ we find:
\bea
&P^2-Z Z_5=\displaystyle{-\left(\frac{\alpha}{\omega} H-\omega\tZ_-\right)^2+\vec P^2-\frac{1}4\left(4\omega^2\tZ_+^2-\left(-\frac{2(1+\alpha)}\omega H  + 2\omega \tZ_-\right)^2\right)}&\nn\\&\displaystyle{=\frac{1+2\alpha}{\omega^2}H^2 +\vec P^2-2 H\tZ_--\omega^2 \tZ_+^2}.&
\eea
In the limit $\omega\to \infty$ we have a divergent term proportional to $\tZ_+^2$. {
Since $\tZ_+$ is a central charge, it is clear that the $\tZ_+^2$ is a Casimir of the NR-VSUSY algebra. In this situation also the finite part is a Casimir and}  
 coincides with the Casimir of the Bargmann algebra \cite{bargmann}
 \be
C_E =\vec P^2-2H\tZ_-.
\ee
In the relativistic case, if the bosonic Casimir vanishes, that is  $P^2-Z Z_5=0$,   an odd Casimir: $G_\mu P^\mu +ZG_5$ 
 \cite{Casalbuoni:2008ez,Casalbuoni:2009en} is also present.
. Using the definitions given in eq. (\ref{eq:8}) and expanding in powers of  $\omega$, we find (we have assumed $Z_+=0$)
\be
G_\mu P^\mu +ZG_5=\frac{\omega^{1/2}}{\sqrt{2}}\left(\vec Q\cdot\vec P+2\tilde Z_-Q_++Q_-H\right)-\frac{\omega^{-3/2}}{\sqrt{2}}(1+2\alpha)Q_+H.
\ee
Let us consider the coefficient of $\sqrt{\omega/2}$  
\be 
C_O=\vec Q\cdot\vec P+2\tilde Z_-Q_++HQ_-.
\ee
It is easily checked that this is a Casimir of the NR-VSUSY if $C_E=0$ (remember that we have assumed  $\tZ_+=0$). Notice also, that
\be
[C_O,C_O]_+=-2\tZ_- C_E.\ee

 As we will see in the next section, the existence of the odd Casimir will imply
that the NR-VSUSY spinning particle shows  a kappa-invariance \cite{Casalbuoni:2008iy}.

{
Notice that the parameter $\alpha$, appearing in the definition of the generators of the contracted algebra in eq. (\ref{eq:8}), does not affect  the NR-VSUSY  algebra. As a consequence, although we do not have a formal proof, we expect that all the expressions depending only on the algebra itself do not depend on the choice of $\alpha$. In the following we will show that the non relativistic limit of the relativistic Lagrangian describing the spinning particle does not depend on $\alpha$.  

\section{Non relativistic limit of the VSUSY particle}\label{sec:1}

In the context of  non-linear realization of  a group symmetry $G$,  the dynamical variables defining the model are nothing but the coset parameters. This allows us to transfer the contraction from the algebra to the dynamical variables, $x^\alpha$, assuming
\be
\sum_\alpha x^\alpha X_\alpha = \sum_{\alpha,\beta} x^\alpha A_\alpha^\beta(\omega)Y_\beta\equiv \sum_\alpha \tilde x^\alpha(\omega)Y_\alpha,
\ee
where we have defined the "contracted" variables 
\be
\tilde x^\alpha=\sum_\beta x^\beta A^\alpha_\beta(\omega).
\ee
The original dynamical model is defined in terms of a lagrangian depending  on  the dynamical variables $x^\alpha$. As a consequence we define the {
lagrangian of the} "contracted" dynamical model as
\be
L_{\rm contracted}(\tilde x)=\lim_{\omega\to \infty} L(x(\tilde x(\omega)).
\ee
The simplest example of this procedure is the non-relativistic limit of the relativistic point particle. One starts from the Poincar\'e symmetry, $IO(1,3)$, in a four-dimensional space-time. The model can be obtained by considering the coset $IO(1,3)/O(1,3)$. The dynamical variables are the  coordinates $x^\mu$, the coset parameters. By performing the contraction to the Galilei group, one obtains the non-relativistic point particle,  {
except for a divergent total derivative.}
The   divergent
term  can be eliminated by introducing the coupling to a U(1) gauge field with vanishing field strength  \cite{Gomis:2000bd}}. {
In the case of the model considered in this paper, this $U(1)$ gauge field is obtained by the gauging of one of the central charges of the VSUSY algebra.}

In this paper we will apply the previous idea to the spinning particle \cite{Barducci:1976qu}. This model is based on the invariance with respect to the VSUSY algebra. The   construction of  the dynamical model as a non-linear realization of V-SUSY has been considered in
\cite{Casalbuoni:2008iy}. {
Here  we will consider  
the contraction of the relativistic model 
to its non-relativistic version, invariant under NR-VSUSY}. In order to get the non-relativistic version, 
we will make use of the contraction defined in the previous Section.

The action for the VSUSY particle, introduced in \cite{Barducci:1976qu,Casalbuoni:2008iy} is given by
\be S\left[x(\tau),\xi(\tau)\right]=\int
d\tau\left(-{\alp}\sqrt{-\left({\dot x}^\mu- i
\xi^\mu{\dot\xi}^5\right)^2}-\beta\left(\dot c +\frac i 2\xi^\mu
\dot\xi_\mu\right)-\gamma\left(\dot {c}_5+\frac i2\xi^5
\dot{\xi^5}\right)\right),\label{eq:action1}\ee
where $x^\mu$ are the space-time coordinates, $\xi^\mu$ is a Grassmann pseudo-vector, $\xi^5$  a Grassmann pseudo-scalar, 
 $c$ and $ c_5$ are the bosonic coordinates associated to the two central charges of the VSUSY algebra , $\tau$ parametrizes the trajectory and 
the parameters $\beta$ and $\gamma$ satisfy
\be
\beta\gamma=-\mu^2,\ee
{
in order the system is invariant under a gauge world-line supersymmetry,
i.e., kappa symmetry
 \cite{Casalbuoni:2008iy}.}
We will make the choice $\beta=-\gamma=-\mu$. The choice $\beta=\mu$ would give rise to divergent terms that are not total derivatives in the NR limit.
The action (\ref{eq:action1}) is Poincar\'e invariant and it has a vector supersymmetry defined by the following variations of the coordinates:
\be \delta x^\mu= i \epsilon^\mu
\xi^5,~~~\delta\xi^\mu=\epsilon^\mu,~~~\delta c=\frac i
2\xi_\mu\epsilon^\mu
\label{eq:19}
\ee 
and 
\be
\delta\xi^5=\epsilon^5,~~~\delta c_5=\frac i2\xi^5\epsilon^5,
\label{eq:20}
\ee
\noindent
where the $\epsilon^\mu, \epsilon^5$ are the supersymmetric
 parameters.
As discussed before, we define the non-relativistic variables starting from the contracted algebra and requiring the following relation (for analogous relations in the bosonic case see \cite{Bergshoeff:2015uaa}):
\be
x^\mu P_\mu+cZ+c_5Z_5+\xi^\mu G_\mu+\xi^5G_5=-tH +\vec x\cdot\vec P +\frac 12(\tc_+ \tZ_+ +\tc_- \tZ_-)+
\txi^0 Q_-+\txi^5 Q_++\txi^i Q_i.\label{eq:28}
\ee
Here we have omitted  the Lorentz group generators, because the related parameters do not enter into the action.
From  {
eq. (\ref{eq:28})} we obtain the following relations 
\bea
&x^0=\displaystyle{\omega  t -\frac{1+\alpha}{2\omega} \tilde c_-},~~~c_-=-\frac{\alpha}{2\omega}\tilde c_-+\omega t,~~~c_+ =\frac 1{2\omega}\tc_,~~{
c_\pm = c\pm c_5},&\nn\\& \xi^0=\displaystyle{\sqrt{\omega\over 2}\txi^5+\sqrt{1\over{2\omega^3}}\txi^0}, ~~~\xi^5=
\sqrt{\omega\over 2}\txi^5-\sqrt{1\over{2\omega^3}}\txi^0,~~~\vec\xi=\sqrt{2\over\omega}\vec\txi.&\label{eq:22}
\eea
The inverse relations are:
\bea
&t= \displaystyle{-\frac 1\omega(\alpha x^0-(1+\alpha)c_-),~~~ \tc_-=-2\omega(x^0-c_-),~~~\tc_+=2\omega c_+},&\nn\\
&\displaystyle{\txi^0=\sqrt{\frac{\omega^3}2}(\xi^0-\xi^5),~~~\txi^5=\frac 1{\sqrt{2\omega}}(\xi^0+\xi^5),~~~
\txi^i=\sqrt{\frac\omega 2}\xi^i}.&
\label{eq:23}
\eea

Then, performing the limit, the  result is
\be
S_{NR}= \int d\tau L_{NR}=\int d\tau\left[\frac 12 M\frac{(\dot{\vec x}-i\vec\txi\dot\txi^5)^2}{\dot t-i\txi^5\dot\txi^5/2}+ i M\vec\txi\cdot\dot{\vec\txi}+\frac M2\frac d{d\tau}(\tilde c_-+i\tilde\xi^0\tilde\xi^5)\right],
\label{eq:24}
\ee
where
\be
M=\frac\mu\omega\ee  {
is assumed to be finite.}
 Notice that there is no divergent term in the non-relativistic expansion. This is due to the presence of the variables   $c_\pm$, associated to the central charges $Z_\pm$,  related to two $U(1)$ curl-free gauge fields. {
 In fact, we can re-express the total derivative terms $\dot c$ and $\dot c_5$ appearing in (\ref{eq:action1}) in the forme $\de_\mu M_i \dot x^\mu,~~i=1,2$.}

In order to get the NR-VSUSY transformations, we notice that the transformations of the NR variables, are obtained, through the eqs. (\ref{eq:23}) from the corresponding transformation laws of the relativistic variables, given in eqs.
(\ref{eq:19}) and (\ref{eq:20}). For infinitesimal transformations, the NR parameters are given by the same combinations defining $\xi^\mu$ and $\xi^5$ in terms of their NR correspondent (see eq. (\ref{eq:22})):
\be
\epsilon^0=\displaystyle{\sqrt{\omega\over 2}\tepsilon^5+\sqrt{1\over{2\omega^3}}\tepsilon^0}, ~~~\epsilon^5=
\sqrt{\omega\over 2}\tepsilon^5-\sqrt{1\over{2\omega^3}}\tepsilon^0,~~~\vec\epsilon=\sqrt{2\over\omega}\vec\tepsilon.
\ee
The  NR variables transform as follows:
\bea
&\delta t=\frac i2 \tepsilon^5\txi^5,~~~\delta \vec x=i\vec\tepsilon\,\txi^5~~~\delta\txi^5=\tepsilon^5,~~~\delta\vec\txi=\vec\tepsilon,&\nn\\&\delta \tc_+=0,~~~\delta \tc_-= -i(\tepsilon^0\txi^5-\tepsilon^5\txi^0),~~~\delta\txi^0= \tepsilon^0,&
\label{eq:26}
\eea
and it is easily seen that $L_{NR}$ is invariant under the transformations (\ref{eq:26}). 
 Eliminating the total derivative term from $L_{NR}$ , this would be quasi-invariant.

Let us now consider the canonical momenta associated to the non relativistic action. We have
\bea
\vec p&=&\frac{\de L_{NR}}{\de\dot{\vec x}}=M\frac{(\dot{\vec x}-i\vec\txi\dot\txi^5)}{\dot t-i\txi^5\dot\txi^5/2},\\
E &=&-\frac{\de L_{NR}}{\de \dot t}=\frac 12 M\frac{(\dot{\vec x}-i\vec\txi\dot\txi^5)^2}{\left(\dot t-i\txi^5\dot\txi^5/2\right)^2},\\
\tilde\pi^5&=&\frac{ \partial_l L_{NR}}{\de \dot\txi^5}= iM\frac{(\dot{\vec x}-i\vec\txi\dot\txi^5)}{\dot t-i\txi^5\dot\txi^5/2}\cdot\vec\txi-\frac i4 M \frac{(\dot{\vec x}-i\vec\txi\dot\txi^5)^2}{\left(\dot t-i\txi^5\dot\txi^5/2\right)^2}\txi^5\nn\\
&=&i\vec p\cdot\vec\txi-\frac i 2 E\txi^5,\\
\vec{\tilde\pi}&=&\frac{\de_l L_{NR}}{\de \dot{\vec\txi}}=-iM\vec\txi.
\label{eq:29}
\eea
Here the derivatives with respect to the Grassmann variables are defined as left derivatives.

These relations imply two first class constraints
\be
\phi=2ME-\vec p^{\,2}=0,~~~~\chi=\tilde\pi^5+\frac i 2 E\txi^5-i\vec p\cdot\vec\txi=0,\ee
and the second class ones
\be
\chi_i=\tilde\pi_i+iM\txi_i=0.\ee
In fact,
\be
\{\chi_i,\chi_j\}=-2iM\delta_{ij},\ee
where we have made use of the following canonical Poisson brackets for the odd variables:
\be
\{\tilde\xi^i,\tilde\pi_j\}=-\delta_j^i,~~~\{\tilde\xi^5,\tilde\pi^5\}=-1.\ee
The second class constraints can be eliminated using the Dirac brackets, 
\be
\{\txi_i,\txi_j\}^*=-i\frac{\delta_{ij}}{2M}, ~~~ \{\txi^5,\tilde\pi^5\}^*=-1.
\ee
{
For the first class constraints we have 
}
\be
\{\chi,\chi\}^*=-iE+i\frac{\vec p^{\, 2}}{2M}=-\frac i {2M}\phi,
\ee
and
\be \{\chi,\phi\}^*=0,\ee
showing that $\phi$ and $\chi$ are indeed first-class constraints.

{
 The {
 symplectic} action of the NR spinning particle is given by
\be\label{canonicalaction}
S_c=\int d\tau\left[-E\dot t+{\vec p }\dot{\vec x}+i M\vec\txi\cdot\dot{\vec\txi}+\tilde\pi^5\dot\txi^5
-\frac e2(2ME-\vec p^{\,2})-
\rho(\tilde\pi^5+\frac i 2 E\txi^5-i\vec p\cdot\vec\txi)\right],
\ee
where the Lagrange multipliers $e, \rho$ multiply the first class constraints.
}

The 
Noether generators of VSUSY are easily found

\be
Q_+=\tilde\pi^5-\frac i 2 H\txi^5,~~~Q_-=0,~~~Q_i={\tilde\pi_i} +ip_i\txi^5-iM\txi_i.
\label{eq:38a}
\ee
Their  Dirac brackets are

\be
\{Q_+,Q_+\}^*=iH,~~~\{Q_+,Q_i\}^*=-ip_i,~~~\{Q_i,Q_j\}^*=2iM\delta_{ij}.
\ee
After quantization one can check that these   Dirac brackets are consistent with the abstract algebra of eq. (\ref{eq:11}), except for the opposite sign in the right hand side. This is due to the fact that the infinitesimal variations in the classical case are generated by $\epsilon G$ rather than by $i\epsilon G$ as implicitly assumed in the case of the abstract algebra. In fact, the charges in eq.  (\ref{eq:38a})  become anti-hermitian after quantization, whereas the abstract charges were supposed to be hermitian. Notice that in this model $Z_-=M$ and $Z_+=0$ and, as a consequence, the NR-VSUSY algebra admits two Casimirs with zero value, corresponding to the two constraints $\phi$ and $\chi$.

\section{Equations of motion}

Since $L_{NR}$ is translationally invariant in time and space, the equations of motion for the bosonic coordinates are 
simply (we will use indifferently $E$ or $H$):
\be
\dot{\vec p}=\dot E=0.
\ee
For the fermionic variables we get
\be
\frac{\de L_{NR}}{\de\dot{\vec\txi}}=-iM\vec\txi,~~~\frac{\de L_{NR}}{\de{\vec\txi}}=iM\dot{\vec\txi}-i\vec p\,\dot\txi^5,
\ee
from which
\be
\dot{\vec\txi}=\frac{\vec p}{2M}\dot\txi^5.
\label{eq:36}
\ee
Recalling the expression (\ref{eq:29}) for $\tilde\pi^5$ we obtain
\be
\frac d{d\tau}\frac{\de L_{NR}}{\de \dot\txi^5}=-\frac i2 E\dot\txi^5+i\vec p\cdot\dot{\vec\txi},~~~
\frac{\de L_{NR}}{\de \txi^5}=\frac i2E\dot\txi^5,
\ee
implying
\be
\dot\txi^5=\frac{\vec p\cdot\dot{\vec\txi}}{E}.
\label{eq:38}
\ee
Notice that multiplying (\ref{eq:36}) by $\vec p$ we get
\be
\vec p\cdot\dot{\vec\txi}=\frac{\vec p^2}{2M}\dot\txi^5.
\ee
This equation is the same as eq. (\ref{eq:38}), after using the constraint $\phi =0$. We see that the equations of motion for the odd variables are not independent, in fact one  is a consequence of the others. The fact that 
 the equations of motions 
 are not independent implies that a local (gauge) symmetry is present in the model.
{
In other words, this implies that a Noether identity is present}. This will be shown explicitly in the next Section.  Also the even constraint generates a relation among the equations of motion. In fact, differentiating the even constraint $\vec p^{\,2}-2EM=0$ we get the identity
\be
\vec p\cdot\dot{\vec p}-2\dot E M=0,\ee
implying that the four bosonic equations, $\dot{\vec p}=\dot E=0$, are not independent. The local symmetry induced by the even constraint is the invariance under reparametrization in the time parameter.

\section{kappa-symmetry}

As we know, the $\phi$-constraint is related to the reparametrization invariance of $L_{NR}$ . Furthermore, the existence of the constraint $\chi$, and the fact that the equations of motion for the fermionic variables are not independent, suggest the existence of a local  (in the time-parameter $\tau$) symmetry. In this Section, we will use  the quantum notation, defining the infinitesimal transformations of a dynamical variable F, as

\be
\delta F=[i\epsilon G,F].
\ee
In the case of the even constraint  the  following local transformation is generated
\be
\delta x_i= - 2\epsilon(\tau) p_i,~~~\delta t(\tau)= -2\epsilon M,\ee
whereas for the odd case:
\be
\delta A=[i\kappa(\tau)\chi,A].
\ee
We get
\be
\delta \vec x=i\vec\xi \kappa,~~~\delta t=\frac i 2 \txi^5\kappa,~~~\delta\vec\txi=\frac{\vec p}{2M}\kappa,~~~
\delta \txi^5=\kappa.
\ee
Let us now consider the variations of the various terms in  $L_{NR}$
\be
\delta(\dot{\vec x}-i\vec\txi\dot\txi^5)=\frac d{d\tau}\left(i\vec\txi\kappa\right)-i\frac{\vec p}{2M}\kappa \dot\txi^5
-i\vec\txi\dot\kappa =i(\dot{\vec\txi}+\frac{\vec p}{2M}\dot\txi^5)\kappa.
\ee
 Then,
\be
\delta\left(\dot t-\frac i2\txi^5\dot\txi^5\right)=\frac d{d\tau}\left(\frac i2\txi^5\kappa\right)-\frac i 2\kappa\dot\txi^5
-\frac i 2\txi^5\dot\kappa=i\dot\txi^5\kappa.\ee

Finally we have to consider the variation of the $\vec\txi$ kinetic term:
\be
\delta(iM\vec\txi\cdot\dot{\vec\txi})=-iM\frac{\vec p}{2M}\cdot\dot{\vec\txi}\,\kappa+iM\frac{\vec\txi\cdot\vec p}{2M}
\dot\kappa.
\ee
The total variation of  $L_{NR}$ can be written as follows
\bea
&&\delta L_{NR}= \vec p\cdot \delta(\dot{\vec x}-i\vec\txi\dot\txi^5)-E\delta\left(\dot t-\frac i2\txi^5\dot\txi^5\right)+\delta(iM\vec\txi\cdot\dot{\vec\txi}) \nn\\&&= i({\vec p}\cdot\dot{ \vec\txi}+\frac{{\vec p}^2}{2M}\dot\txi^5)\kappa-iE\dot\txi^5\kappa-i\frac{\vec p}{2}\cdot\dot{\vec\txi}\,\kappa+i\frac{\vec\txi\cdot\vec p}{2}\dot\kappa.
\eea
Adding the previous variations we  finally get
\be
\delta L_{NR}=\frac d{d \tau}\left(\frac i 2\vec p\cdot{\vec\txi}\,\kappa\right),
\ee
where we have used  the fact that $\phi$ is identically zero, since here both $E$ and $\vec p$ should be considered as functions of the lagrangian variables.

Using the $\kappa$-symmetry. we can fix $\txi^5$ to zero. In this way the constraint $\chi$ becomes second class. The lagrangian simplifies to:

\be
L_{NR}=\frac 12 M\frac{\dot{\vec x}^{\,2}}{\dot t}+ i M\vec\txi\cdot\dot{\vec\txi},
\ee
where we have omitted the total derivative appearing in (\ref{eq:24}). In this form, the related action is still invariant under reparametrization of the parameter $\tau$. We can choose the gauge $t=\tau$, making also the constraint $\phi$ second class, obtaining
\be
L_{NR}=\frac 12 M{\dot{\vec x}^{\,2}}+ i M\vec\txi\cdot\dot{\vec\txi}.
\ee

\section{The spinning particle in a Newton-Cartan metric}

{
In this Section we will study the non-relativistic limit of the spinning particle in a 
{
torsionless}
Newton-Cartan 
background \cite{Cartan:1923zea}.  We will introduce the gravitational interaction of the relativistic spinning particle
 by means of a set of vierbein fields, $E_\mu^A$, where  the index $A$ refers to the flat target space-time,, $A=(0,\hi), \hi=1,2,3$, whereas the index $\mu=(0,i), i=1,2,3$ refers to the curved target space-time. 
Then, we define the "contracted" vierbeins in terms of the same linear transformation $A_\alpha^\beta(\omega)$ defining the contracted dynamical variables \cite{Bergshoeff:2015uaa}. On the other hand, we leave the dynamical variables unchanged. In this second way of proceeding we get a dynamical model interacting with a gravitational field appearing as the contraction of the original one, in such a way to preserve the contraction of the original symmetry.         }

The Newton-Cartan metric is defined in terms of the temporal and spatial vielbeins $(\tau_\mu, e_\mu^{\hat i})$, where $ {\hat i}=1,2,3$ is an index in the flat target space, and $\mu$ is defined as before  \cite{Kuchar:1980tw,
Andringa:2010it,Andringa:2012uz}. The vielbeins satisfy the following relations
\bea
&e_\mu^{\hi}e^\mu_{\hj}=\delta_{\hj}^{\hi}, ~~~e_\mu^{\hi}e^\nu_{\hi}=\delta_\mu^\nu -\tau_\mu\tau^\nu,~~~\tau_\mu\tau^\mu =1,&\cr
&\tau_\mu e^\mu_{\hi}=\tau^\mu e_\mu^{\hi}=0.&
\eea
In order to reduce the generic background metric to the Newton-Cartan one, we will make use of the contraction of the VSUSY algebra as defined in eq. (\ref{eq:8}). In particular, we notice that the expression for $H$ in eq. (\ref{eq:9}), implies a mixing between the Poincar\'e generators and the central charge $Z_-$
\be
H= -\frac\omega 2(2P_0+Z_-).\ee
Therefore, besides introducing   a gravitational field, we  should  also introduce a gauge field associated with the $U(1)$ symmetry generated by $Z_-$. On the other hand, since the Poincar\'e group has no central extension (contrarily to the Galilei group), this new field should be non-dynamical, and therefore  with zero curvature \cite{Gomis:2000bd,Bergshoeff:2015uaa}.

{
We perform the contraction of the vierbeins  assuming  the following correspondendence} 
\bea
&x^0\to E_\mu^0, ~~~c_-\to M_\mu,&\cr
&t\to \tau_\mu,~~~\tilde c_-\to\tilde m_\mu,&
\eea
{
where $M_\mu$ is the $U(1)$ gauge field associated to $Z_-$.}
 Then, using eqs. (\ref{eq:22}) and (\ref{eq:23}), it follows at once
 \bea
 &\dd{E_\mu^0=\omega \tau_\mu-\frac{1+\alpha}{2\omega}\tilde m_\mu,~~~M_\mu = \omega\tau_\mu-\frac \alpha{2\omega}\tilde m_\mu},&\cr
 &\dd{\tau_\mu=-\frac 1\omega (\alpha E_\mu^0-(1+\alpha)M_\mu),~~~\tilde m_\mu =2\omega(M_\mu-E_\mu^0)}.&
 \label{eq:62}
 \eea
 {
 Notice that we do not introduce  gauge fields associated to the
 fermionic generators $G_\mu, G_5$ and $ Q_\pm, Q_i$.}
The change in the contraction procedure, going from the flat to the curved case, does not affect the Grassmann variables. As a consequence we will define the corresponding contracted variables, exactly as in eqs. (\ref{eq:22}) and (\ref{eq:23}), by taking the non-contracted variables $(\xi^A,\xi^5)$ in the flat target-space.

The coupling of a 
background gravitational field with the spinning particle has been studied in \cite{Barducci:1976wc}. In the present notations the Lagrangian is given by
\bea
L&=&-\mu\sqrt{-\eta_{AB}(E_\mu^A\dot x^\mu-i\xi^A\dot\xi^5)(E_\mu^B\dot x^\mu-i\xi^B\dot\xi^5)}\cr
&+&\frac i 2 \mu\left[\eta_{AB}\xi^A\dot\xi^B+
\xi^A \omega_{\mu  [AB]}\dot x^\mu\xi^B\right]-\frac i 2\xi^5\dot\xi^5+\mu M_\mu\dot x^\mu,
\label{eq:63}
\eea
where $\omega_{\mu [AB}]$ is the relativistic spin connection.
Since $M_\mu$ has zero curvature, the expression $M_\mu\dot x^\mu$ is a total $\tau$-derivative. In fact, $M_\mu\dot x^\mu$ can be identified with $\mu^2\dot c_-(x(\tau))$.

Except for the part containing the spin connection,  the NR limit obtained by sending $\omega$ to infinity, proceeds exactly as in the flat case, with the correspondence $\dot x^0\to E_\mu^0\dot x^\mu$, $\dot x^i\to e_\mu^{\hi}\dot x^\mu$, and $\dot c_-\to M_\mu\dot x^\mu$, and taking into account eq. (\ref{eq:62}).
 Notice also that the quadratic divergence arising from $\mu E_\mu^0\dot x^\mu$ (remember that $\mu = M\omega$), is cancelled by the first term in the expression of $M_\mu$ given in eq. (\ref{eq:62}). The result we find is
\bea
 L_{NR}- L_{NR}^{conn}&=&\frac 12 M\frac{(e_\mu^{\hi}\dot x^\mu-i\txi^{\,\hi}\dot\txi^5)^2}{\tau_\mu\dot x^\mu-i\txi^5\dot\txi^5/2}+ i M\vec\txi\cdot\dot{\vec\txi}+i\frac M2\frac d{d\tau}(\tilde\xi^0\tilde\xi^5)+\frac 12 M\tilde m_\mu\dot x^\mu,
\eea
where $L_{NR}^{conn}$ is the NR limit of the part of the Lagrangian relative to the spin connection.

Let us now {
consider the part of the lagrangian that contains} the spin connection. We define the following one-forms
\be
E^A=E_\mu^A dx^\mu,~~~~ \omega_B^A=\omega_{\mu B}^Adx^\mu,\ee
where the one-form defining the spin connection, $ \omega_B^A$, can be evaluated using the first Cartan structure equation
\be
dE^A-\omega_{~B}^A\theta^B=0\label{eq:73}.\ee
{
The expression for the spin connection is given explicitly in 
\cite{Freedman:2012zz}. 
The result is
\be
\omega_{\mu [A B]} = \omega_{\mu [\nu\rho]} E_A^\nu E_B^\rho,\ee
where $E_B^\nu$ is the  inverse vierbein, defined by
\be
E_A^\mu E_\mu^B=\eta_A^B\ee
and
\be \omega_{\mu [\nu\rho]}=\frac 12 \left[\Omega_{[\mu\nu]\rho}-\Omega_{[\nu\rho]\mu}+\Omega_{[\rho\mu]\nu}\right]E_A^\nu E_B^\rho,\ee
with
\be
\Omega_{[\mu\nu]\rho}=(\de_\mu E_\nu^C-\de_\nu E_\mu^C)E_{C\rho}.\ee
It is convenient to define the quantities
\bea
&&\alpha_{\mu[AB]}=\Omega_{[\mu\nu]\rho}E_A^\nu E_B^\rho,\cr
&&\beta_{\mu[AB]}=\Omega_{[\nu\rho]\mu}E_A^\nu E_B^\rho,\cr
&&\gamma_{\mu[AB]}=\Omega_{[\rho\mu]\nu}E_A^\nu E_B^\rho.
\eea
It follows
\be
\omega_{\mu [A B]}=\frac 12 \left[\alpha_{\mu[AB]}-\beta_{\mu[AB]}+\gamma_{\mu[AB]}\right].\label{eq:81}
\ee
Let us evaluate the quantities in parenthesis. We begin with the first and the third term
\be
\alpha_{\mu [AB]}= (\de_\mu E_{B\nu}-\de_\nu E_{B\mu})E_A^\nu,
\ee
\be
\gamma_{\mu [AB]} = (\de_\rho E_{A\mu}-\de_\mu E_{A\rho})E^\rho_B =-\alpha_{\mu[BA]},
\ee
\be
\beta_{\mu[AB]}= (\de_\nu E_\rho^C-\de_\rho E_\nu^C)E_{C\mu}E_A^\nu E_B^\rho.\ee
In order to evaluate the limit of this expression, we make use of the first two eqs. in (\ref{eq:62}).  Recalling that $M_\mu$ has zero curvature, we find
\be
(\tau_{\mu.\nu}-\tau_{\nu,\mu})=\frac \alpha {2\omega^2} (\tilde m_{\mu,\nu}-\tilde m_{\nu,\mu}),\ee
therefore  $(\tau_{\mu.\nu}-\tau_{\nu,\mu})$ goes as $1/\omega^2$ when $\omega\to\infty$. {
This is equivalent to say that in the NR limit the spinning particle should be  coupled to a torsionless NC background. }
The expressions for the inverse vierbeins, at the order in $1/\omega$ we are interested in are
\be
E_0^\mu=\frac 1 \omega \tau^\mu+\frac{(1+\alpha)}{2\omega^3}\tau^\mu\tau^\rho \tilde m_\rho +\left({\cal O}(\omega^{-5})\right),\ee
\be
E^\mu_{\hi}=e^\mu_{\hi}+\frac{(1+\alpha)}{2\omega^2}\tau^\mu e^\rho_{\hi}\tilde m_\rho
+\left({\cal O}(\omega^{-4})\right).\ee
In the NR limit we have
\be
\alpha_{\mu[0\hi]} =\frac 1\omega \tau^\nu(\de_\mu e_{\hi\nu}-\de_\nu e_{\hi\mu})\equiv \frac 1\omega\alpha_{\mu[0\hi]}^{NC},\ee
\be
\alpha_{\mu[\hi 0]}= \frac 1{2\omega} e^\nu_{\hi}(\de_\mu \tilde m_\nu-\de_\nu\tilde m _\mu)\equiv \frac 1\omega\alpha_{\mu[\hi 0]}^{NC},
\ee
\be
\alpha_{\mu[\hi \hj]}= (\de_\mu e_{\hi\nu}-\de_\nu e_{\hi\mu})e_{\hj}^\nu,
\ee
\be
\beta_{\mu [0\hi]} = \frac 1\omega\tau^\nu e^\rho_{\hi}e_{\hj\mu}(\de_\nu e_\rho^{\hj}-\de_\rho e^{\hj}_\nu) +{\cal O}\left(\omega^{-2}\right) \equiv \frac 1\omega\beta_{\mu [0\hi]}^{NC},
\ee
\be 
\beta_{\mu [\hi 0]}=-\beta_{\mu [0\hi]},
\ee
\be
 \beta_{\mu [\hi \hj]}=e_{\hi}^\nu e_{\hj}^\rho \left[e_{\hat k\mu}(\de_\nu e_\rho^{\hat k}-\de_\rho e_\nu^{\hat k})-\frac 12\tau_\mu(\de_\nu \tilde m_\rho-\de_\rho\tilde m_\nu)\right].
  \ee
The superscript $NC$, that we have introduced here. refers to the Newton-Cartan quantities.

The result for the spin connection is
\be
\omega_{\mu[0\hi]}=-\omega_{\mu[\hi 0]} =\frac 1{2\omega} (\alpha_{\mu[0\hi]}^{NC}-\beta_{\mu[0 \hi]}^{NC}-\alpha_{\mu[\hi 0]}^{NC})\equiv\frac 1\omega \omega_{\mu 0\hi]}^{NC},
\ee
\be
\omega_{\mu[\hi\hj]}=\frac 12 (\alpha_{\mu[\hi\hj]}^{NC}-\beta_{\mu[\hi\hj]}^{NC}-\alpha_{\mu[\hj\hi]}^{NC})\equiv \omega_{\mu[\hi\hj}]^{NC},
\ee
with
\be\label{spinconnection}
\omega_{\mu [0\hi]}^{NC}=\frac 12\left[\tau^\nu(\de_\mu e_{\hi\nu}-\de_\nu e_{\hi\mu})-\tau^\nu e^\rho_{\hi}e_{\hj\mu}(\de_\nu e_\rho^{\hj}-\de_\rho e^{\hj}_\nu) -\frac 1{2} e^\nu_{\hi}(\de_\mu \tilde m_\nu-\de_\nu\tilde m _\mu)\right],
\ee
\bea\label{spinconnection1}
\omega_{\mu[\hi\hj]}^{NC}&=&\frac 12 \Big[ (\de_\mu e_{\hi\nu}-\de_\nu e_{\hi\mu})e_{\hj}^\nu-e_{\hi}^\nu e_{\hj}^\rho \left[e_{\hat k\mu}(\de_\nu e_\rho^{\hat k}-\de_\rho e_\nu^{\hat k})-\frac 12\tau_\mu(\de_\nu \tilde m_\rho-\de_\rho\tilde m_\nu)\right]\cr &-& (\de_\mu e_{\hj\nu}-\de_\nu e_{\hj\mu})e_{\hi}^\nu\Big].
\eea
As anticipated, also for the Newton-Cartan connections we get a result independent of $\alpha$ 
and agree with the ones obtained in \cite{Andringa:2010it}. 

%
%
%
%
Keeping in mind that $\omega_{\mu[\hi 0]}$ is multiplied by $\mu \xi^{\hi}\xi^0$, which is given by
\be
\mu \xi^{\hi}\xi^0=\omega M \sqrt{\frac 2 \omega}\tilde\xi^{\hi}\sqrt{\frac\omega 2}\tilde\xi^5=\omega M\xi^{\hi}\tilde\xi^5\label{eq:96},\ee
we see that we get a finite result. The same happens for the rotation part of the connection, since
\be
\mu \xi^{\hi}\xi^{\hj}= 2 M \tilde\xi^{\hi}\xi^{\hj}\label{eq:97}.\ee
Therefore the result of the NR limit for the spin connection contribution to the Lagrangian is
\bea
 L_{NR}^{conn}=iM\txi^5\txi^{\hi}\dot x^\mu\omega_{\mu[0\hi]}^{NC}+
  iM\txi^{\hi}\txi^{\hj}\dot x^\mu\omega_{\mu[\hi\hj]}^{NC}.
 \eea
The presence of the spin connection relative to the Galilei boost should not surprise. In fact, the variable $\tilde \xi^5$ arising from a linear combination of $\xi^0$ and $\xi^5$ does not transform trivially under a Galilei boost. 
Let us define
\be
A_\mu = iM\txi^5\txi^{\hi}\omega_{\mu[0\hi]}^{NC}+
iM\txi^{\hi}\txi^{\hj}\omega_{\mu[\hi\hj]}^{NC}+\frac 12 M\tilde m_\mu.\label{eq:107}
\ee
Then, we have (neglecting the total derivative)
\bea
L_{NR}&=&\frac 12 M\frac{(e_\mu^{\hi}\dot x^\mu-i\txi^{\,\hi}\dot\txi^5)^2}{\tau_\mu\dot x^\mu-i\txi^5\dot\txi^5/2}+ i M\vec\txi\cdot\dot{\vec\txi}+A_\mu\dot x^\mu,\label{eq:100}
\eea
{
 that is the lagrangian of a non-relativistic spinning particle in Newton-Cartan
background.}
\section{Equations of motion in the Newton-Cartan case}
Let us evaluate the momenta:
\be
p_\mu=\frac{\de L_{NR}}{\de\dot x^\mu}= M\frac{(e_\nu^{\hi}\dot x^\nu-i\txi^{\,\hi}\dot\txi^5)}{\tau_\nu\dot x^\nu-i\txi^5\dot\txi^5/2}e_\mu^{\hi}-\frac 12 M\frac{(e_\nu^{\hi}\dot x^\nu-i\txi^{\,\hi}\dot\txi^5)^2}{(\tau_\nu\dot x^\nu-i\txi^5\dot\txi^5/2)^2}\tau_\mu+A_\mu.
\ee
Defining
\be
{\cal P}_\mu=p_\mu-A_\mu,
\ee
and projecting along the vielbeins, we get 

\be
{\cal P}_\mu\tau^\mu =-\frac 12 M\frac{(e_\nu^{\hi}\dot x^\nu-i\txi^{\,\hi}\dot\txi^5)^2}{(\tau_\nu\dot x^\nu-i\txi^5\dot\txi^5/2)^2}
\ee
and
\be
{\cal P}_\mu e^\mu_{\hi}=M\frac{(e_\nu^{\hi}\dot x^\nu-i\txi^{\,\hi}\dot\txi^5)}{\tau_\nu\dot x^\nu-i\txi^5\dot\txi^5/2}.
\ee
From which we get the 
{
mass-shell constraint
}
\be\label{massshellNC}
\phi=2M{\cal P}_\mu\tau^\mu+({\cal P}_\mu e^\mu_{\hi})^2=0
\ee
{
and the odd constraint
}
\be\label{kappaNC}
\chi=\tilde\pi^5-i{\cal P}_\mu e^\mu_{\hi}\xi^{\hi}-\frac i 2{\cal P}_\mu\tau^\mu\tilde\xi^5=0.
\ee
These two constraints are the analogs of the constraints 
we found in the flat case.

{
It is useful to introduce the "curved NR" Grassmann variables} 
\be
\lambda^\mu=\tilde{e}^\mu_A\zeta^A,~~~ {
\tilde{e}^\mu_A=(\tau^\mu, e^\mu_{\hi})},
\ee
where 
\be{
\zeta^A=(\frac 12 \tilde\xi^5, \tilde\xi^{\hi}), ~~~A=( 0,\hi).}\ee
 Their Dirac bracket are given
by
\bea
&\{\zeta^A,\zeta^B\}^*=-\frac i{2M}\sum_i\delta_i^A\delta_i^B,&\nn\\
&\{\lambda^\mu,\lambda^\nu\}^* =-\frac{i}{2  M}\sum_ie^\mu_i e^\nu_i=-\frac{i}{2  M} h^{\mu\nu}=-\frac{i}{2 M}(\eta^{\mu\nu}-\tau^\mu\tau^\nu).&
\eea
The odd constraint becomes
\be
\chi=\tilde\pi^5-i{\cal P}_\mu\lambda^\mu.
\ee
{
In the same notations, we can write
\be
A_\mu=iM\zeta^A\zeta^B\omega_{\mu[AB]}^{NC}+\frac 12 M\tilde m_\mu.\ee}

Now we should compute the Dirac bracket of the constraints $\phi, \chi$; for this it is useful to compute 
\be
\{{\cal P}_\mu,{\cal P}_\nu\}^* = M \left( (R_{\mu\nu}^{NC})_{AB} S^{AB}+\frac 1 2\left(\frac{\de\tilde m_\mu}{\de x^\nu}-\frac{\de\tilde m_\nu}{\de x^\mu}\right)\right), 
\ee 
where

\be
(R_{\mu\nu}^{NC})_{AB}=\frac {\de\omega_{\nu[AB]}^{NC}}{\de x^\mu} -\frac {\de\omega_{\mu[AB]}^{NC}}{\de x^\nu} -\omega_{\nu[A\hi]}^{NC}\omega_{\mu[\hi B]}^{NC}+ \omega_{\mu[A\hi]}^{NC}\omega_{\nu[\hi B]}^{NC}\ee
is the curvature tensor for the NC structure and
\be 
S^{AB}=i\zeta^A\zeta^B\label{eq:115}\ee
are the spin generators. Furthermore
\be
\{{\cal P}_\mu,\lambda^\nu\}^* =\lambda^\rho \Gamma^\nu_{\mu\rho},
\ee
where $  \Gamma^\nu_{\mu\rho}$ are the Christoffel symbols associated to the NC structure given by
 \bea
\Gamma^\nu_{\mu\rho}=-\tilde e^A_\rho(\de_\mu\tilde e_A^\nu+\omega_{\mu[A\hi]}^{NC}\tilde e^\nu_{\hi}),
 \eea
{
which agrees with connection obtained in 
\cite{Andringa:2010it}. In this paper it was proved that these connections are symmetric in the lower indices}

Using the previous  results we  obtain 
\bea
\{\chi, \chi\}^*&=& \frac i{2M}\phi +2{\cal P}_\mu\Gamma^\mu_{\rho\nu}\lambda^\rho
\lambda^\nu+\frac 14 M(\de_\mu\tilde m_\nu-\de_\nu\tilde m_\mu)\tau^\mu e_{\hi}^\nu\tilde \xi^{\hi}\tilde\xi^5\cr
 &&-\frac M2[(R_{\mu\nu}^{NC})_{AB}S^{AB}
 ]
 [(\tau^\mu e^\nu_k-\tau^\nu e^\mu_k)\tilde\xi^5\tilde\xi^{\hat k}+(e^\mu_ke^\nu_\ell-e^\mu_\ell e^\nu_k)\tilde\xi^k\tilde\xi^\ell].\label{eq:118}
 \eea
{
To preserve the continuity with the  case of a flat background, it is important to require that  the two constraints $(\phi,\chi)$ are still first class. To this end, let us notice that the connection $\Gamma^\mu_{\rho\nu}$ is symmetric in the lower indices. 
Being saturated with the antisymmetric quantity $\lambda^\rho\lambda^\nu$ gives zero contribution.The term proportional to the curvature is zero due to the Bianchi identity. However, the third term  proportional to the curl of $\tilde m_\mu$ does not vanish. Therefore we must require that the vector field associated to the $U(1)$ symmetry is curl-free. We will make use of this condition from now on.
} 
  
  Therefore
\be
\{\chi, \chi\}^*= \frac i{2M}\phi\label{eq:125}.
\ee
Since the constraint $\chi$ is odd,  the following Jacobi identity 
\be
\{\chi,\{\chi,\chi\}^*\}^*+\{\chi,\{\chi,\chi\}^*\}^*+\{\chi,\{\chi,\chi\}^*\}^*=0\ee
is not trivial. Using  eq. (\ref{eq:125}) it follows
\be
\{\chi,\phi\}^*=0.
\ee
Therefore  the two constraints $(\phi, \chi)$ are first class not only in the flat case, but also in a torsionless NC background,
with a $U(1)$ connection with zero field strength.
Correspondingly the term in the non-relativistic lagrangian containing the gauge field is a total derivative.

Since  the constraint $\chi$ implies the existence  of the kappa-symetry, we see that a requirement of kappa-symmetry gives informations about background. In the case of superbranes in a supergravity background this
interplay among the world volume symmetry and  a supergravity background  implies the on-shell equations of motion of  supergravity \cite{Grisaru:1985fv}.}

 The Dirac hamiltonian can be written as
\be
 H=\alpha(2M{\cal P}_\mu\tau^\mu+({\cal P}_\mu e^\mu_{\hi})^2)+
 \beta(\tilde\pi^5-i{\cal P}_\mu e^\mu_{\hi}\xi^{\hi}-
 \frac i 2{\cal P}_\mu\tau^\mu\tilde\xi^5).
 \ee
In order to get  the equations of motion for the space-time coordinates, Grassmann and spin variables it is useful to consider the  gauge $\beta=0$. In this case we have 
 \bea
 \dot x^\mu&=& 2\alpha(M\tau^\mu+e^\mu_{\hi} e^\nu_{\hi} {\cal P}_\nu),\cr
\dot\txi^5&=&0, \cr
\dot\txi^{\hi}&+&(\omega_{\mu[\hi\hj]}^{NC}\xi^{\hj}-\frac 12 \omega_{\mu[0\hi]}^{NC}\txi^5)\dot x^\mu=0,\cr
\dot{\cal P}_\mu&-&\left(e_{\hi}^\rho\frac{\de e^{\hi}_\nu}{\de x^\mu}+\tau^\rho\frac{\de \tau_\nu}{\de x^\mu}\right){\cal P}_\rho\dot x^\nu=
M  (R_{\mu\nu}^{NC})_{AB} S^{AB}\dot x^\nu.
\eea
We can also have the equations of motion of the spin variables $S^{AB}$ defined in eq. (\ref{eq:115}).
They are given by
\be
\dot S^{\hi 0}+\omega_{\mu[\hi\hj]}S^{\hj 0}\dot x^\mu=0, ~~~\dot S^{\hi\hj}+\omega_{\mu[\hi A]}S^{A\hj}\dot x^\mu-\omega_{\mu[\hj A]}S^{A\hi}\dot x^\mu=0.
\ee
Keeping in mind that in this gauge $\zeta^0=\tilde\xi^5/2$ is a constant of motion, this is equivalent to say that the covariant derivative of the spin generators vanishes.
In order to get the second order equations for the ${x^{\mu}}'s$ variables we must express
$ {\cal P}^\nu$ in terms of $\dot x^\nu$. The final  
result is 
\be\label{papapetru}
\ddot x^\nu+\Gamma^\nu_{\rho\mu} \dot x^\rho \dot x^\mu-\frac{\dot N}{ N}\dot x^\nu=Mh^{\nu\rho}
(R_{\rho\mu}^{NC})_{AB} S^{AB}\dot x^\mu,
\ee
where $N=\tau_\mu\dot  x^\mu$.\footnote{An analogous equation for the coordinates was found in 
\cite{Duval:1976ht}. However the evolution of the spin is different.} Choosing the gauge with $\alpha=$  constant it follows, from  $\tau_\mu\dot x^\mu=2\alpha M$, that $\dot N$ vanishes. Notice also that the connection is given by
\bea\label{connection1}
\Gamma^\nu_{\rho\mu}&=&
 \frac 12 [h^{\nu\sigma}(\frac{\partial h_{\sigma\mu}}{\partial x^\rho}+
 \frac{\partial h_{\sigma\rho}}{\partial x^\mu}-\frac{\partial h_{\mu\rho}}{\partial x^\sigma})+
 \tau^\nu( \frac{\partial \tau_{\rho}}{\partial x^\mu}{ +} \frac{\partial \tau_{\mu}}{\partial x^\rho})].
\eea
 By putting the Grassmann variables to zero, the  equation (\ref{papapetru}) reduces to the geodesic equation for a scalar particle in a torsionless NC background \cite{Kuchar:1980tw}. 
It is interesting to notice that in a generic Newton-Cartan background, the left-hand side of eq. ({\ref{papapetru}) would contain a  term with three velocities $\dot x^\mu$ given by :
 \be
 \frac 12 h^{\nu\mu}h_{\rho\lambda}\left(\de_\mu\tau_\sigma-\de_\sigma\tau_\mu\right) \dot x^\rho \dot x^\lambda\dot x^\sigma\ee
 However, such a term vanishes in a torsionless 
 Newton-Cartan background as it is in our case.}

\section{Conclusions and Outlook}

In this paper we have constructed the action of a non-relativistic spinning particle moving in a general torsionless Newton-Cartan background. The spinning particle is described in terms of Grassmann variables.
The model has two gauge symmetries, diffeomorphism and kappa symmetry. {
The invariance under kappa symmetry implies 
that the gauge field associated to one of the central extensions of the VSUSY algebra 
\cite{Casalbuoni:2008iy} has vanishing field strength. }

The equations for the space-time coordinates
do not follow the geodesic equations, instead the motion is governed by the non-relativistic analog of  the Papapetrou equation \cite{Papapetrou:1951pa} with a coupling of the spin to the NC curvature.  

One could study the construction of a non-relativistic superparticle in a  NC background by null reduction of the analogous relativistic spinning particle in one dimension more.

\vskip3mm\noindent
{\bf Note added in proof:}
After having completed this work we received the paper \cite{Kluson:2017pzr} where the behaviour of a scalar particle and of a supersymmetric particle in a 3-dimensional NC background were examined. 
Despite of some similarity  this approach and ours are very different, as well as the model considered here.

\vskip3mm
\acknowledgments
We acknowledge discussions with Paul Townsend, Jorge Zanelli
and Christian Duval and Peter Horvathy for pointing us \cite{Duval:1976ht}.
JG has been supported  by FPA2013-46570-C2-1-P and Consolider CPAN and by
the Spanish goverment (MINECO/FEDER) under project MDM-2014-0369 of ICCUB (Unidad de Excelencia Mar\'\i a de Maeztu).

government


\begin{thebibliography}{37}
\bibitem{review}
S.~Sachdev, 
 ``Quantum Phase
Transitions," Cambridge University Press (2011), ISBN - 978-0-521-51468-2.
 \bibitem{review1}
 Y. ~Liu, K. ~Schalm, Y.-W. ~Sun and J.
~Zaanen, ``Holographic Duality in Condensed Matter Physics,"  Cambridge University Press (2015), ISBN: 9781107080089.

\bibitem{Son:2008ye}
  D.~T.~Son,
  ``Toward an AdS/cold atoms correspondence: A Geometric realization of the Schrodinger symmetry,''
  Phys.\ Rev.\ D {\bf 78} (2008) 046003
  doi:10.1103/PhysRevD.78.046003
  [arXiv:0804.3972 [hep-th]].

\bibitem{Balasubramanian:2008dm}
  K.~Balasubramanian and J.~McGreevy,
  ``Gravity duals for non-relativistic CFTs,''
  Phys.\ Rev.\ Lett.\  {\bf 101} (2008) 061601
  doi:10.1103/PhysRevLett.101.061601
  [arXiv:0804.4053 [hep-th]].

\bibitem{Herzog:2008wg}
  C.~P.~Herzog, M.~Rangamani and S.~F.~Ross,
  ``Heating up Galilean holography,''
  JHEP {\bf 0811} (2008) 080
  doi:10.1088/1126-6708/2008/11/080
  [arXiv:0807.1099 [hep-th]].

\bibitem{Kachru:2008yh}
  S.~Kachru, X.~Liu and M.~Mulligan,
  ``Gravity duals of Lifshitz-like fixed points,''
  Phys.\ Rev.\ D {\bf 78} (2008) 106005
  doi:10.1103/PhysRevD.78.106005
  [arXiv:0808.1725 [hep-th]].
  
  
\bibitem{Son:2013rqa}
  D.~T.~Son,
  ``Newton-Cartan Geometry and the Quantum Hall Effect,''
  arXiv:1306.0638 [cond-mat.mes-hall].
  
\bibitem{Geracie:2016bkg}
  M.~Geracie,
  ``Galilean Geometry in Condensed Matter Systems,''
  arXiv:1611.01198 [hep-th].
  
\bibitem{Janiszewski:2012nb}
  S.~Janiszewski and A.~Karch,
  ``Non-relativistic holography from Horava gravity,''
  JHEP {\bf 1302} (2013) 123
  doi:10.1007/JHEP02(2013)123
  [arXiv:1211.0005 [hep-th]].

\bibitem{Wu:2014dha}
  C.~Wu and S.~F.~Wu,
  ``Horava-Lifshitz gravity and effective theory of the fractional quantum Hall effect,''
  JHEP {\bf 1501} (2015) 120
  doi:10.1007/JHEP01(2015)120
  [arXiv:1409.1178 [hep-th]].

\bibitem{Cartan:1923zea}
  E.~Cartan,
  ``Sur les vari\'et\'es \`a connexion affine et la th\'eorie de la relativit\'e g\'en\'eralis\'ee. (premi\`ere partie),''
  Annales Sci.\ Ecole Norm.\ Sup.\  {\bf 40} (1923) 325.
 
\bibitem{Horava:2009uw}
  P.~Horava,
  ``Quantum Gravity at a Lifshitz Point,''
  Phys.\ Rev.\ D {\bf 79} (2009) 084008
  doi:10.1103/PhysRevD.79.084008
  [arXiv:0901.3775 [hep-th]].
  
\bibitem{Kuchar:1980tw}
  K.~Kuchar,
  ``Gravitation, Geometry, And Nonrelativistic Quantum Theory,''
  Phys.\ Rev.\ D {\bf 22} (1980) 1285.
  doi:10.1103/PhysRevD.22.1285
  
\bibitem{Andringa:2012uz} 
  R.~Andringa, E.~Bergshoeff, J.~Gomis and M.~de Roo,
  ``'Stringy' Newton-Cartan Gravity,''
  Class.\ Quant.\ Grav.\  {\bf 29}, 235020 (2012)
  doi:10.1088/0264-9381/29/23/235020
  [arXiv:1206.5176 [hep-th]].
    
  
  
  
  
\bibitem{Jensen:2014aia}
  K.~Jensen,
  ``On the coupling of Galilean-invariant field theories to curved spacetime,''
  arXiv:1408.6855 [hep-th].

\bibitem{Hartong:2014pma}
  J.~Hartong, E.~Kiritsis and N.~A.~Obers,
  Phys.\ Rev.\ D {\bf 92} (2015) 066003
  doi:10.1103/PhysRevD.92.066003
  [arXiv:1409.1522 [hep-th]].
    
    
\bibitem{Papapetrou:1951pa}
  A.~Papapetrou,
  ``Spinning test particles in general relativity. 1.,''
  Proc.\ Roy.\ Soc.\ Lond.\ A {\bf 209} (1951) 248.
  doi:10.1098/rspa.1951.0200
  
 \bibitem{Casalbuoni:2008iy}
R.~Casalbuoni, J.~Gomis, K.~Kamimura  and G.~Longhi, {{Space-time vector
  supersymmetry and massive spinning particle}}, JHEP {\bf 02} (2008)
  \href{http://dx.doi.org/10.1088/1126-6708/2008/02/094}{094},
\href{http://arxiv.org/abs/0801.2702}{{\tt arXiv:0801.2702 [hep-th]}}
   
   \bibitem{Barducci:1976qu}
A.~Barducci, R.~Casalbuoni  and L.~Lusanna, {{"Supersymmetries and the
  pseudoclassical relativistic electron"}}, Nuovo Cim. {\bf A35} (1976)
\href{http://dx.doi.org/10.1007/BF02730291}{377}


  
  \bibitem{deAzcarraga:1982dw}
J.~A.~de Azc\'arraga and J.~Lukierski, ``Supersymmetric Particles
With Internal Symmetries And Central Charges,'' Phys.\ Lett.\ B {\bf
113} (1982) 170.
J.~A.~de Azc\'arraga and J.~Lukierski, ``Supersymmetric Particles In
N=2 Superspace: Phase Space Variables And Hamiltonian Dynamics,''
Phys.\ Rev.\ D {\bf 28} (1983) 1337.
\bibitem{Siegel:1983hh}
W.~Siegel, ``Hidden Local Supersymmetry In The Supersymmetric
Particle Action,'' Phys.\ Lett.\ B {\bf 128} (1983) 397.


  
\bibitem{LevyLeblond:1967zz}
  J.~M.~Levy-Leblond,
  ``Nonrelativistic particles and wave equations,''
  Commun.\ Math.\ Phys.\  {\bf 6} (1967) 286.
  doi:10.1007/BF01646020
  
  
\bibitem{Gomis:1985uf}
  J.~Gomis and M.~Novell,
  ``A Pseudoclassical Description for a Nonrelativistic Spinning Particle. 1. The Levy-leblond Equation,''
  Phys.\ Rev.\ D {\bf 33} (1986) 2212.
  doi:10.1103/PhysRevD.33.2212
  
\bibitem{Barducci:1976wc} 
  A.~Barducci, R.~Casalbuoni and L.~Lusanna,
  ``Classical Spinning Particles Interacting with External Gravitational Fields,''
  Nucl.\ Phys.\ B {\bf 124}, 521 (1977).
  doi:10.1016/0550-3213(77)90419-9



  
\bibitem{Bergshoeff:2015uaa}
  E.~Bergshoeff, J.~Rosseel and T.~Zojer,
  ``Newton-Cartan (super)gravity as a non-relativistic limit,''
  Class.\ Quant.\ Grav.\  {\bf 32} (2015) no.20,  205003
  doi:10.1088/0264-9381/32/20/205003
  [arXiv:1505.02095 [hep-th]].
  
\bibitem{Gomis:2000bd}
  J.~Gomis and H.~Ooguri,
  ``Nonrelativistic closed string theory,''
  J.\ Math.\ Phys.\  {\bf 42} (2001) 3127
  doi:10.1063/1.1372697
  [hep-th/0009181].
  
\bibitem{Gauntlett:1989qe}
  J.~P.~Gauntlett, K.~Itoh and P.~K.~Townsend,
  ``Superparticle With Extrinsic Curvature,''
  Phys.\ Lett.\ B {\bf 238} (1990) 65.
  doi:10.1016/0370-2693(90)92101-N


\bibitem{Bagger:1996wp}
  J.~Bagger and A.~Galperin,
  ``A New Goldstone multiplet for partially broken supersymmetry,''
  Phys.\ Rev.\ D {\bf 55} (1997) 1091
  doi:10.1103/PhysRevD.55.1091
  [hep-th/9608177].

\bibitem{Gauntlett:1990nk}
  J.~P.~Gauntlett, J.~Gomis and P.~K.~Townsend,
  ``Particle Actions as {Wess-Zumino} Terms for Space-time (Super)symmetry Groups,''
  Phys.\ Lett.\ B {\bf 249} (1990) 255.
  doi:10.1016/0370-2693(90)91251-6
     
 
  
   
\bibitem{Casalbuoni:2008ez}
  R.~Casalbuoni, F.~Elmetti, J.~Gomis, K.~Kamimura, L.~Tamassia and A.~Van Proeyen,
  ``Vector Supersymmetry: Casimir operators and contraction from OSp(3,2|2),''
  JHEP {\bf 0901} (2009) 035
  doi:10.1088/1126-6708/2009/01/035
  [arXiv:0812.1982 [hep-th]].


\bibitem{Casalbuoni:2009en} 
  R.~Casalbuoni, F.~Elmetti, J.~Gomis, K.~Kamimura, L.~Tamassia and A.~Van Proeyen,
  ``Vector Supersymmetry from OSp(3,2|2): Casimir Operators,''
  Fortsch.\ Phys.\  {\bf 57}, 521 (2009)
  doi:10.1002/prop.200900016
  [arXiv:0901.4862 [hep-th]].

  





   \bibitem{bargmann}
 V.~Bargmann,
{``On Unitary Ray Representations Of Continuous Groups,''}
  Annals Math.\  {\bf 59} (1954) 1.
  
\bibitem{Andringa:2010it}
  R.~Andringa, E.~Bergshoeff, S.~Panda and M.~de Roo,
  ``Newtonian Gravity and the Bargmann Algebra,''
  Class.\ Quant.\ Grav.\  {\bf 28} (2011) 105011
  doi:10.1088/0264-9381/28/10/105011
  [arXiv:1011.1145 [hep-th]].
  
\bibitem{Freedman:2012zz}
  D.~Z.~Freedman and A.~Van Proeyen,
  ``Supergravity'', Cambridge University Press 2012.
  
\bibitem{Grisaru:1985fv}
  M.~T.~Grisaru, P.~S.~Howe, L.~Mezincescu, B.~Nilsson and P.~K.~Townsend,
  ``N=2 Superstrings in a Supergravity Background,''
  Phys.\ Lett.\  {\bf 162B} (1985) 116.
  doi:10.1016/0370-2693(85)91071-8

  
   
\bibitem{Kluson:2017pzr}
  J.~Kluson,
  ``Canonical Analysis of Non-Relativistic Particle and Superparticle,''
  arXiv:1709.09405 [hep-th].
  
\bibitem{Duval:1976ht}
  C.~Duval and H.~P.~Kunzle,
  Rept.\ Math.\ Phys.\  {\bf 13} (1978) 351.
  doi:10.1016/0034-4877(78)90063-0
  
   
  
  
    
\end{thebibliography}
\end{document}